\documentstyle[twoside,ijmpc1]{article}
\def\qed{\hbox{${\vcenter{\vbox{            
   \hrule height 0.4pt\hbox{\vrule width 0.4pt height 6pt
   \kern5pt\vrule width 0.4pt}\hrule height 0.4pt}}}$}}

\renewcommand{\thefootnote}{\fnsymbol{footnote}}    

\def\bsc{{\sc a\kern-6.4pt\sc a\kern-6.4pt\sc a}}   
\def\bflatex{\bf L\kern-.30em\raise.3ex\hbox{\bsc}\kern-.14em
T\kern-.1667em\lower.7ex\hbox{E}\kern-.125em X}

\begin{document}
\setlength{\textheight}{7.7truein}  

\thispagestyle{empty}

\markboth{\protect{\footnotesize\it Integrability of a
Non-autonomous
 Coupled KdV System}}
{\protect{\footnotesize\it Ay\c{s}e Karasu (Kalkanl{\i})
  and Tuba K{\i}l{\i}\c{c}}}

\normalsize\textlineskip

\setcounter{page}{1}

\copyrightheading{}         

\vspace*{0.88truein}

\centerline{\bf INTEGRABILITY OF A NON-AUTONOMOUS}
\vspace*{0.035truein}
\centerline{\bf COUPLED KdV SYSTEM}

\vspace*{0.37truein}
\centerline{\footnotesize AY\c{S}E KARASU(KALKANLI)\footnote{
E-mail: akarasu@metu.edu.tr}
and TUBA KILI\c{C}}
\baselineskip=12pt
\centerline{\footnotesize\it Department of Physics, Middle East
Technical University}
\baselineskip=10pt
\centerline{\footnotesize\it  06531 Ankara,
Turkey}

\vspace*{15pt}          
\vspace*{0.225truein}
\publisher{(received date)}{(revised date)}

\vspace*{0.25truein}
\abstracts{The
Painlev\'{e} property of  coupled, non-autonomous Korteweg-de
Vries (KdV) type of systems is studied. The conditions under which
the systems pass the Painlev\'{e} test for integrability are
obtained. For some of the integrable cases, exact solutions are
given.
}{}{}

\vspace*{5pt}
\keywords{non-autonomous KdV systems; Painlev\'{e} analysis;
exact solutions}


\vspace*{1pt}\textlineskip  
\section{Introduction}     
\vspace*{-0.5pt}
\noindent
For a better
understanding of complicated physical phenomena scientists have
experienced that it is necessary to introduce mathematical models
whose time evolutions might show some features very similar to
those of the original phenomena. These models are usually systems
of nonlinear differential equations. These equations can be solved
by the use of approximation techniques. But the range of
applicability and usefulness of these solutions increase the
interest on the exact solutions and on the solution generating
methods for nonlinear equations. Before attempting to solve an
equation one usually needs to know whether the equation is
integrable or not. The difficulty of obtaining solutions by use of
inverse scattering transform technique, makes this information
valuable. One of the powerful tools to obtain this information is
the Painlev\'{e} test for integrability. There are strong
evidences that all integrable equations have Painlev\'{e}
property, that is, all solutions are single valued around movable
singularities.$^{1}$
\eject
In this work, we consider the non-autonomous, coupled KdV type systems
\begin{eqnarray}
u_t&=&u_{xxx}+ a(t) uu_x,\nonumber\\ v_t&=&v_{xxx}+
b(t) (uv)_x,
\end{eqnarray}
where $a(t)$ and $b(t)$ are some arbitrary
functions. We apply the Painlev\'{e} test for integrability to
system (1), following the Weiss-Kruskal algorithm of singularity
analysis $^{2,3}$ and obtain the conditions on the functions $a(t)$
and $b(t)$. We find the subclasses of these equations that possess
the Painlev\'{e} property. By using the truncated expansions we
obtain the exact solutions for some of these subclasses,
explicitly.


\setcounter{footnote}{0}
\renewcommand{\thefootnote}{\alph{footnote}}

\section{The Non-autonomous Coupled KdV type Systems}
\noindent
Following
the approach of Weiss {\it et al} $^2$, we assume that the
solutions of (1) can be represented by the expansions, \begin{eqnarray}
u(x,t)&=& \sum_{r=0}^{\infty}u_r(x,t)\phi^{r+\alpha},\nonumber\\
v(x,t)&=& \sum_{r=0}^{\infty}v_r(x,t)\phi^{r+\beta}, \end{eqnarray} where
$\alpha$ and $\beta$ are integers, $u_r(x,t)$ and $v_r(x,t)$ are
analytic functions in a neighborhood of the singularity manifold
$\phi(x,t)$. A hypersurface $\phi(x,t)=0$ is noncharacteristic of
the system (1) if $\phi_t \phi_x \neq 0$. We choose$^3$
$\phi(x,t)=x+\psi(t)$, without loss of generality, hence the
coefficients $(u_r,v_r)$ are independent of $x$. This is the
simplest choice for the test, but it cannot be used to obtain the
particular solutions. The substitution of $u(x,t)=
\sum_{r=0}^{n}u_r(t)\phi^{r+\alpha},\;\;\; v(x,t)=
\sum_{r=0}^{n}v_r(t)\phi^{r+\beta}$ into (1) determines the
branches, i.e. the admissible dominant behavior of solutions, and
the corresponding positions $r$ of the resonances where the
arbitrary functions can appear in the expansions (2). The leading
order analysis gives that \begin{equation} \alpha=-2 ,\;\;\;
u_0=-\frac{12}{a(t)} ,\;\;\;
(\beta-2)\left[\beta(\beta-1)-12\frac{b(t)}{a(t)}\right]v_0(t)=0.
\end{equation} The branches satisfying (3) are \begin{equation}
\beta_1 = 2 ,\;\; \beta_2
= -m ,\;\; \beta_3 = m+1 \end{equation}
where $m$ is a non-negative integer
and \begin{equation}
b(t) = \frac{a(t)}{12}(m^2+m). \end{equation} \\ \\ For each branch,
the corresponding positions $r$ of resonances are \begin{eqnarray} \beta_1& =
&2, \;\; r = 0, m-1, -(m+2),\nonumber\\ \beta_2& = &-m, \;\; r =
0, m+2, 2m+1,\nonumber\\ \beta_3& = &m+1, \;\; r = 0, 1-m,
-(2m+1). \end{eqnarray} For every branch there exists the common resonance
$r=0$. For $\beta_1$ and $\beta_3$, at least one of the resonances
always stands in a negative position. The second branch, $\beta_2
= -m$, has two positive resonances for every value of $m$. Hence,
the second branch is  generic: the expansions (2) with (3)
represent the general solutions near  singularity. Next, we find
from (1) the recursion relations for the coefficients $u_r(t) \;
(r=0,1,2,\ldots)$ of the expansions (2). We see that the
resonances occur at $r=-1,4,6$. The resonance at $r=-1$
corresponds to the arbitrariness of $\phi(x,t)$. For the other
resonances $(r=4,6)$ the recursion relations turn out to be
consistent if \begin{equation} a_{tt}(t)a(t)-3a_t^2(t)=0. \end{equation}
On the other
hand, the recursion relations for $v_r(t)$ are depend on $m$. For
each value of $m$, there exist different recursion relations with
different resonances, but $r=0$ is common for all those cases. We
check every case up to $m=15$ and find the following cases: \begin{eqnarray}
m&=&0 , \;\; r=-1,0,1,2,4,6;\nonumber\\ m&=&2 , \;\;
r=-1,0,4,4,5,6;\nonumber\\ m&=&3 , \;\; r=-1,0,4,5,6,7; \end{eqnarray}
for which, the compatibility conditions are automatically satisfied
for each resonances and the system (1) passes the Painlev\'{e}
test if $a(t)$ satisfies (7). We find that equation (7) has the
solution \begin{equation}
a(t)=\pm[-2c_1t+2c_2]^{-1/2}, \end{equation} where $c_1$ and
$c_2$ are integration constants. It follows that the system (1)
possesses the Painlev\'{e} property if $a(t)= k$ and
$a(t)=\frac{k}{\sqrt{t}}$, where $k$ is any non-zero constant.

For $m=3$, it is obvious from (5) that $a(t)=b(t)$ in the system
(1). The case $a(t)=b(t)=\frac{k}{\sqrt{t}}$ corresponds to the
perturbation system of the cylindrical KdV (cKdV) equation.$^4$
For the other case,
$a(t)=b(t)=k$, the system (1) is the perturbative KdV system and
is studied in Ref. 5. In the following sections we study two of the
systems (1), corresponding to $m=2$, in detail.

\subsection{Jordan KdV Systems}
\noindent
We consider the system of equations
\begin{eqnarray}
u_t&=&u_{xxx}+ 2kuu_x,\nonumber\\
v_t&=&v_{xxx}+ k (uv)_x,
\end{eqnarray}
which corresponds to $m=2$ and passes the Painlev\'{e} test.
Actually, this system of equations is known as a Jordan KdV system
and was studied in Refs. 6 and 7.
To gain more information on (10), we define the transformations
by truncating the series expansions (2) on constant level as follows:
\begin{equation}
u=\frac{u_0}{\phi^2}+\frac{u_1}{\phi}+u_2, \;\;\;\;\;
v=\frac{v_0}{\phi^2}+\frac{v_1}{\phi}+v_2,
\end{equation}
where $u_2(x,t)$ and $v_2(x,t)$ satisfy equations (10) and can be
choosen as  $u_2(x,t)=0$ and $v_2(x,t)=0$. Inserting  above expressions
for $u$ and $v$ into system (10) and setting the coefficients of
each power of $\phi$ to zero, we have
\begin{equation}
u_0=-\frac{6}{k}\phi_x^2, \;\;\;\;\;\;\; u_1=\frac{6}{k}\phi_{xx},
\end{equation}
\begin{eqnarray}
\phi_t-4\phi_{xxx}+3\frac{\phi_{xx}^2}{\phi_x}&=&0,\\
\phi_{xxxx}-2\frac{\phi_{xx}\phi_{xxx}}{\phi_x}+\frac{\phi_{xx}^3}
{\phi_x^2}&=&0, \end{eqnarray}
\begin{eqnarray}
v_1=-\frac{v_{0,x}}{\phi_x}+\frac{\phi_{xx}}{\phi_x^2}v_0,\\
v_{0,xx}-3\frac{\phi_{xx}}{\phi_x}v_{0,x}-2(\frac{\phi_{xxx}}
{\phi_x}-2\frac{\phi_{xx}^2}{\phi_x^2})v_0&=&0,\\
v_{0,t}-(2\frac{\phi_{xxx}}{\phi_x}-\frac{\phi_{xx}^2}{\phi_x^2})
v_{0,x}&=&0. \end{eqnarray}
By introducing $\psi(x,t)$, such that
$\phi_x=\psi^2$, equations (13) and (14) can be written as \begin{eqnarray}
\psi_t-4\psi_{xxx}&=&0,\\
\left(\frac{\psi_{xx}}{\psi}\right)_x&=&0. \end{eqnarray}
These equations
have solutions of the form
\begin{equation}
\psi(x,t)= c_1 e^{[\beta(t)+ \alpha(t)(x-1)]}+c_2 e^{-[\beta(t)+
\alpha(t)(x-1)]}, \end{equation} and
\begin{equation}
\psi(x,t)=c_1 x+c_2 \end{equation} where $\beta(t)= \int \alpha(t) dt$, $c_1$
and $c_2$ are constants.

As a special case, we can choose $\alpha(t)=\alpha=constant$ and
$c_1=c_2=1$, so that the solution (20) can be written as
\begin{equation}
\psi(x,t)=2\cosh \alpha \theta, \end{equation} where $\theta(x,t)=x+4\alpha^2
t$. The corresponding solution for $\phi(x,t)$ is,
\begin{equation}
\phi(x,t)=\frac{1}{\alpha}\sinh 2\alpha \theta +
2\theta + 16\alpha^2 t.
\end{equation}
Using this solution we see that equations (16) and (17) are satisfied
if
\begin{equation}
v_{0,t}-4\alpha^2 v_{0,x}=0.
\end{equation}
Then,
\begin{equation}
v_0(x,t)=f(x+4\alpha^2t)=f(\theta), \end{equation} where $f$ is an arbitrary
function of its argument. Now, using (12), (15), (23) and (25) in
(11) we obtain the exact solutions of Jordan KdV system (10),
\begin{eqnarray}
u(x,t)&=&-\frac{24}{k}\left[\left(\frac{1+\cosh 2\alpha \theta}
       {\frac{1}{\alpha}\sinh 2\alpha \theta +
       2\theta + 16\alpha^2 t }\right)^2
       -\frac{\alpha \sinh 2\alpha \theta}
       {\frac{1}{\alpha}\sinh 2\alpha \theta +
       2\theta + 16\alpha^2 t}\right],\nonumber \\ \nonumber\\
v(x,t)&=&\frac{f(\theta)}{(\frac{1}{\alpha}\sinh 2\alpha \theta +
       2\theta + 16\alpha^2 t)^2}
       -\frac{\left[\frac{f^{\prime}
       (\theta)}{2(1+\cosh 2\alpha \theta)}-\frac{\alpha f(\theta)
       \sinh 2\alpha \theta}{(1+\cosh2 \alpha \theta)^2}\right]}
       {(\frac{1}{\alpha}\sinh 2\alpha \theta +
       2\theta + 16\alpha^2 t)}.
\end{eqnarray}
These functions can be plotted by using $Mathematica$.$^8$
Some results are given in $Figures(1)$ and $(2)$.

Next, we consider the solution in (21) and find the expression for
$\phi(x,t)$ as
\begin{equation}
\phi(x,t)= c_1^2 \left(\frac{x^3}{3}-4t\right)+ c_1 c_2 x^2 +
c_2^2 x \end{equation} that leads to the rational solutions \begin{eqnarray}
u(x,t)&=&-\frac{6}{k}\left(\frac{\phi_x^2}{\phi^2}-
\frac{\phi_{xx}}{\phi}\right)^2,\nonumber\\ v(x,t)&=& d_1
\left(\frac{\phi_x^2}{\phi^2}- \frac{\phi_{xx}}{\phi}\right)+ d_2
\left(
\frac{\phi_x^{3/2}}{\phi^2}-\frac{\phi_{xx}\phi_x^{-1/2}}{2\phi}\right)
\end{eqnarray} of Jordan KdV system (10) where $d_1$ and $d_2$ are arbitrary
constants.

\subsection{Non-autonomous Jordan KdV Systems}
\noindent
As a
second example, we consider the system of equations \begin{eqnarray}
u_t&=&u_{xxx}+ \frac{2}{\sqrt {t}} uu_x,\nonumber\\ v_t&=&v_{xxx}+
\frac{1}{\sqrt {t}} (uv)_x, \end{eqnarray} which corresponds to the case
$m=2$ in (8) and passes the Painlev\'{e} test. This system of
equations is known as non-autonomous Jordan KdV system and is
given in Ref. 9. Inserting the expansions
\begin{equation}
u=\frac{u_0}{\phi^2}+\frac{u_1}{\phi}, \;\;\;\;\;
v=\frac{v_0}{\phi^2}+\frac{v_1}{\phi}, \end{equation} into (29) and setting
the coefficients of each power of $\phi$ to zero, we obtain
\begin{equation}
u_0=-6\sqrt{t}\phi_x^2, \;\;\;\;\;\;\;  u_1= 6\sqrt{t}\phi_{xx},
\end{equation}
\begin{eqnarray}
\phi_t-4\phi_{xxx}+3\frac{\phi_{xx}^2}{\phi_x}&=&0,\\
\phi_{xxxx}-2\frac{\phi_{xx}\phi_{xxx}}{\phi_x}+\frac{\phi_{xx}^3}
{\phi_x^2}+\frac{\phi_x}{6t}&=&0, \end{eqnarray}
\begin{eqnarray}
v_1=-\frac{v_{0,x}}{\phi_x}+\frac{\phi_{xx}}{\phi_x^2}v_0,\\
v_{0,xx}-3\frac{\phi_{xx}}{\phi_x}v_{0,x}-2(\frac{\phi_{xxx}}
{\phi_x}-2\frac{\phi_{xx}^2}{\phi_x^2})v_0&=&0,\\
v_{0,t}-(2\frac{\phi_{xxx}}{\phi_x}-\frac{\phi_{xx}^2}{\phi_x^2})
v_{0,x}+\frac{5}{6t}v_0&=&0. \end{eqnarray}
The equations (32) and (33) are
compatible, i.e.$(\phi_{xxxx})_t = (\phi_t)_{xxxx}$, and can be
solved by the substitution $\phi_x=\psi^2$, where \begin{eqnarray}
\psi_t-4\psi_{xxx}&=&0,\nonumber\\ \psi_{xx}+\left[\frac{x}{12
t}+\alpha(t)\right]\psi&=&0. \end{eqnarray}
However, the last equation can
only be solved  in terms of Airy functions.$^{10}$ The result is
\begin{equation}
\psi(x,t)=(1/t)^{1/3}\left[c_1 Ai(z)+ c_2 Bi(z)\right] \end{equation}
where
\begin{equation}
z(x,t)=\frac{[(-1/t)^{1/3}(x+12c_0)]}{2^{2/3}3^{1/3}}, \;\;\;\;
\alpha(t)= \frac{c_0}{t} \end{equation} and $c_0$, $c_1$, $c_2$ are constants.
The corresponding solution for $\phi(x,t)$ is \begin{eqnarray}
\phi(x,t)&=&(1/t)^{2/3}(12c_0+x)[c_1 Ai(z)+ c_2 Bi(z)]^2
\nonumber\\ &+&
2^{2/3} 3^{1/3}(1/t)^{1/3}[c_1
Ai^{\prime}(z)+ c_2 Bi^{\prime}(z)]^2. \end{eqnarray}
Then,
\begin{eqnarray}
u&=& 6\sqrt{t} (\ln
{\phi})_{xx} \nonumber\\ v&=&
\frac{v_0}{\phi^2}-\frac{1}{\phi}\left(\frac{v_0}{\phi_x}\right)_x,
\end{eqnarray} are the exact solutions of the system of equations (29) if
(35) and (36) are satisfied. Note that, these equations  are
linear in $v_0$ and a particular solution is $v_0=C\sqrt{t}
\phi_x^2$ with  $C=constant$. In this particular case, $v$ is
proportional to $u$, i.e. $v=-(C/6)u$. This implies that the
system of equations in (29) reduces to a cKdV equation after the
transformation $u \rightarrow \sqrt{t}$ $u$.$^{11-13}$ In Ref.13,
a hierarchy of solutions for the cKdV equation is derived in terms
of Airy functions. These solutions can be obtained from (41)
together with (40).

\nonumsection{Acknowledgements}
\noindent
We are grateful to Dr. Sergei Sakovich for his guidence
and useful comments. We also acknowledge the referee for his
careful reading and suggestions on the manuscript.
This work is supported in part by
the Scientific and Technical Research Council of Turkey (TUBITAK).

\nonumsection{References}

\end{document}